\documentstyle[sprocl,epsfig]{article}
\input{psfig}
\def\be{\begin{equation}}
\def\ee{\end{equation}}
\def\bea{\begin{eqnarray}}
\def\eea{\end{eqnarray}}

\begin{document}
\title{KINEMATICAL DEPENDENCE OF NLO CORRECTIONS TO SEMI-INCLUSIVE SPIN DEPENDENT OBSERVABLES.}
\author{{ D.DE FLORIAN,  C.A.GARCIA CANAL}}
\address{
Laboratorio de F\'{\i}sica Te\'{o}rica,
Departamento de F\'{\i}sica \\
Universidad Nacional de La Plata,  C.C. 67 - 1900 La Plata - 
Argentina}
\author{
 { S.JOFFILY}}
\address{ 
 Centro Brasilero de Pesquisas Fisicas \\ Rua Xavier Sigaud 150, Urca, 22.290.180 Rio de Janeiro - Brazil}
\author{  { R.SASSOT} }
\address{Departamento de F\'{\i}sica, 
Universidad de Buenos Aires \\ 
Ciudad Universitaria, Pab.1 
(1428) Bs.As. - 
Argentina}
\maketitle

\abstracts{We analyse the phenomenological consequences of target and current fragmentation
contributions at next to leading order in semi-inclusive spin dependent deep
inelastic scattering asymmetries.}

\section{Introduction:}

Recently, there has been an increasing interest in spin dependent semi-inclusive deep inelastic processes motivated by the inminence of high precision semi-inclusive experiments\cite{1,27}, and by theoretical considerations related to facto\-ri\-za\-tion\cite{2,7}. Both, open the possibility to extract information about the spin of the proton and study hadronization phenomena.

Originally, the observables proposed to be measured, the so called spin dependent asymmetries, were thought in terms of its QCD leading order (LO)\cite{21} descomposition which implied a trivial or eventually vanishing dependence on the kinematical variable $z$: the hadron energy fraction. The experiments proposed so far are designed to measure this observables in certain specific range of this variable.
As the naive description does not take into account events coming from the target fragmentation, which dominate at low  values of $z$, the low $z$-cut is chosen is such a way that most of these events are discarded. In order to increase the statistics, the data are then integrated  over the measured range assuming the trivial LO $z$-dependence of the observable.
In this picture, once the target fragmentation region is eliminated,
both the $z$ dependence of the asymmetry in the restricted interval and that of the
integrated observable coming from the  $z$-cut, become trivial.

However, two facts spoil this simple picture, target fragmentation is always present and NLO corrections introduce a non trivial dependence\cite{2,7}. It is then highly desirable to have an estimate of how these target fragmentation effects are suppresed by the kinematical cut and how dramatic is the residual next to leading order (NLO) $z$ dependence. These are the main objectives of this presentation.

\section{Semi-Inclusive Asymmetries:}

It is customary to define semi-inclusive spin asymmetries $A_{1\,N}^{h}$, as the ratio between
the differences of semi-inclusive deep inelastic cross sections for the production of a hadron $h$ off a target $N$ with opposite helicities ($\Delta \sigma^h_N$) and the correspondent unpolarized cross section ($\sigma^h_N$), and difference asymmetries $A_{N}^{h^+-h^-}$, for the production of hadrons with opposite charges as
\begin{equation}
A_{1\,N}^{h}=\frac{Y_M}{\lambda Y_P}\frac{\Delta \sigma^h_{N}}{\sigma^h_{N}}
\ \ \ \ \ \ \ A_{N}^{h^+-h^-}=\frac{Y_M}{\lambda Y_P}\frac{\Delta \sigma^{h^+}_{N}-\Delta \sigma^{h^-}_{N}}
{\sigma^{h^+}_{N}-\sigma^{h^-}_{N}}
\end{equation}
where $Y_M, \, Y_P$ are kinematical factors and $\lambda$ is the helicity of the incoming lepton.
In the naive parton model these asymmetries reduce to expressions like 
\begin{equation}
A_{1\,N}^{h}=\frac{\sum_{i} e^2_{i}\Delta q(x) D_{h/i}(z)}{\sum_{i} e^2_{i} q(x) D_{h/i}(z)}
\end{equation}
for single asymmetries, or
\begin{eqnarray}
A_{D}^{\pi^+-\pi^-}=\frac{\Delta u_v(x) +\Delta d_v(x)}{u_v(x)+d_v(x)}
 \frac{(D_u^{\pi^+}(z)-D_d^{\pi^+}(z))}{(D_u^{\pi^+}(z)-D_d^{\pi^+}(z))}, \nonumber \\ A_{p}^{\pi^+-\pi^-}=\frac{4 \Delta u_v(x) -\Delta d_v(x)}{4 u_v(x)-d_v(x)} \frac{(D_u^{\pi^+}(z)-D_d^{\pi^+}(z))}{(D_u^{\pi^+}(z)-D_d^{\pi^+}(z))}
\end{eqnarray}
for pion production on deuterium and proton targets respectively.
In the first case, eq.2, the $z$ dependence is that of the already known unpolarized fragmentation functions and for difference asymmetries, eq.3, this dependence cancels.
In the experimental determination of these asymmetries what are actually measured are the cross sections integrated over a range of the variable $z$, and eventually the data are analised taking into account the LO dependence.   

This simple picture for the asymmetries is obiously spoiled when either target fragmentation effects or next to leading order corrections are included.
The latter  implies that the factors in eq.2 have to be replaced by NLO convolutions that also  prevents the cancelation of the $z$ dependence  in eq.3. Adding to this, at NLO target fragmentation effects must be necesarily included  using fracture functions 
for a consistent factorization os collinear divergences\cite{2}. The full expressions for NLO cross sections can be found in reference\cite{2,18,19}.

\section{Results:}

In this section we compute semi-inclusive spin asymmetries for proton and deuteron targets including NLO QCD corrections and contributions coming from the target fragmentation region. For these we use  NLO fragmentation functions\cite{5}, parton distributions\cite{4,3} and a model for spin dependent fracture functions\cite{7}.

In figures (1a) and (1b) we show the naive (solid lines) and the NLO corrected (dashed lines) single asymmetry results for the production of positive hadrons off protons imposing two different kinematical cuts ({\bf a)} $z>0.20$ as in the COMPASS proposal\cite{6},  and 
{\bf b.)} $z>0.10$ as in that of HERMES\cite{27}). For comparison we also include data from EMC\cite{9} and SMC\cite{1} experiments and the errors expected by COMPASS\cite{6}. 

In these asymmetries the size of the corrections, which is not significant regarding  the precision of the available data, are dominated by NLO current fragmentation effects,
fracture functions do not contribute significantly.  These results have been obtained using polarized parton distributions with non negligible gluon polarization. As can be expected, for sets without gluon polarization the size of the corrections are even smaller.
Identical results are obtained for the production of negative hadrons and for
neutron and deuteron targets. 
 
\setlength{\unitlength}{1.mm}
\begin{figure}[hbt]
\begin{picture}(120,48)(0,0)
\put(0,0){\mbox{\epsfxsize5.3cm\epsffile{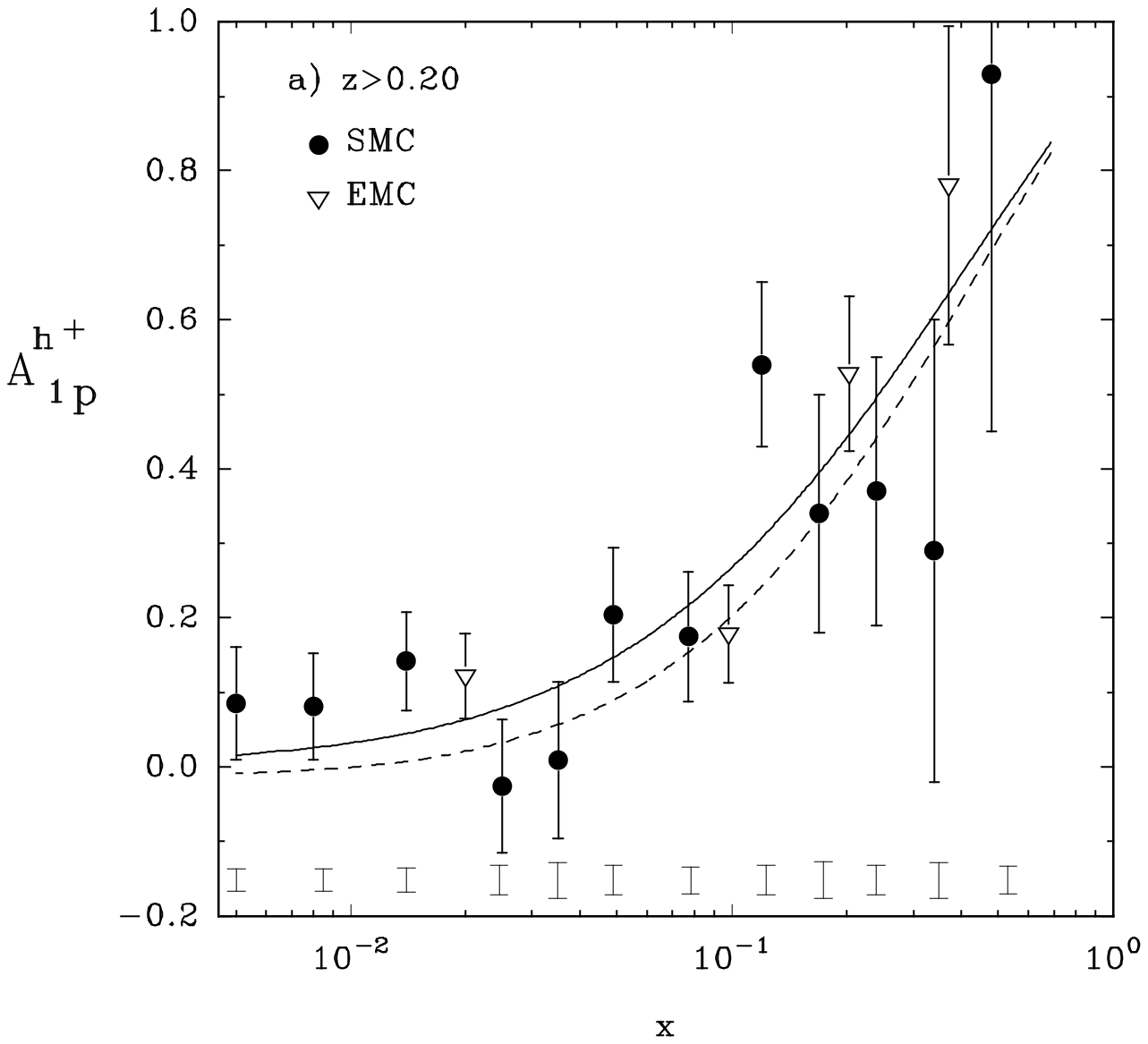}}}
\put(60,0){\mbox{\epsfxsize5.3cm\epsffile{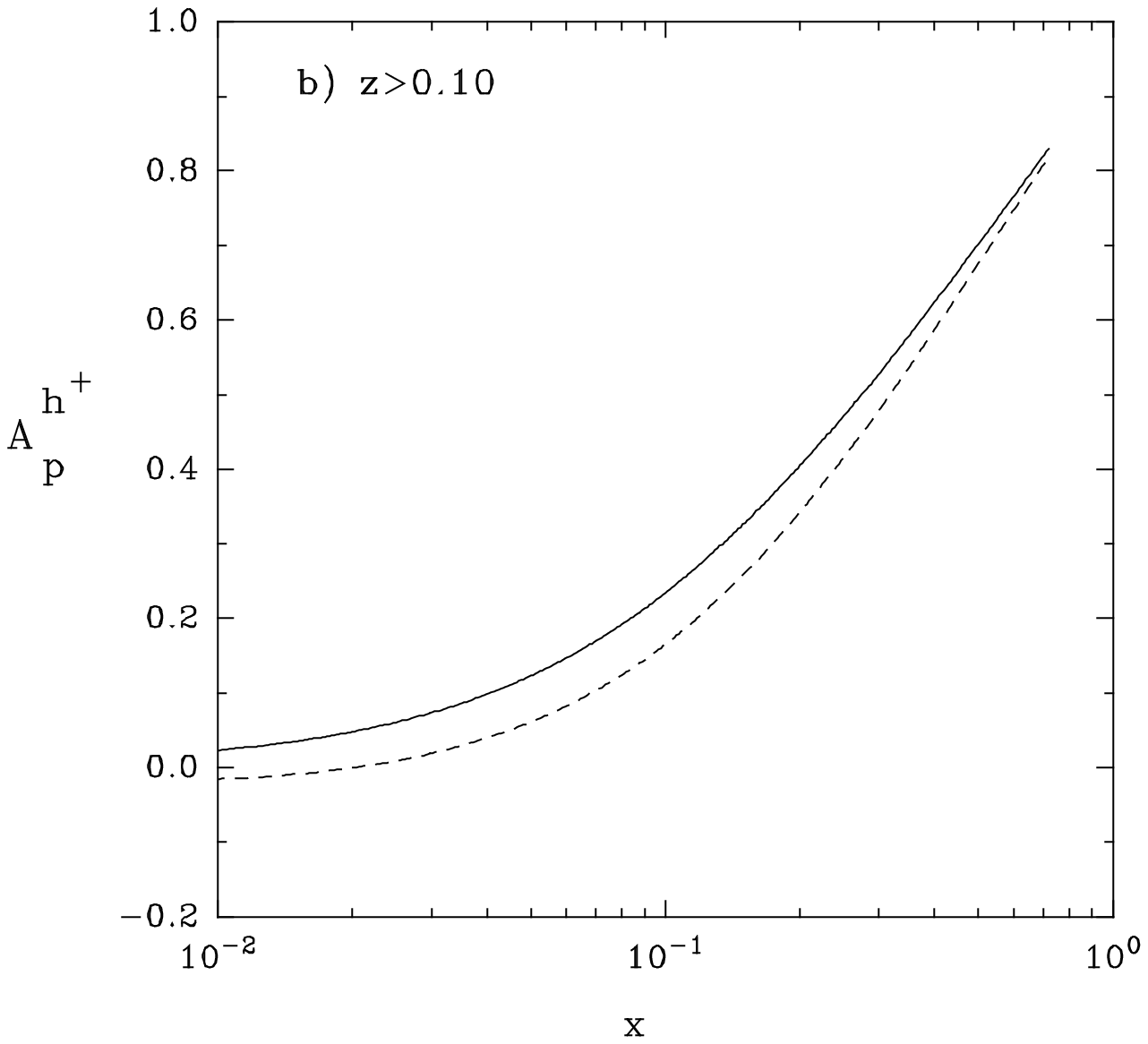}}}
\end{picture}
\caption{Semi-inclusive asymmetries for muoproduction of charged pions and kaons on a proton target with:  {\bf a)}  $z>0.2$, {\bf b)}  $z>0.1$
\label{fig:radish}}
\end{figure}

These figures show also a very mild $z$-cut dependence which can be exploited for increasing the statistics within on experiment, or just for a safe comparison of the future experimental results.  

In figures (2a) and (2b), we show the naive (solid lines) and the NLO corrected (dashed lines) results for difference asymmetries imposing two different kinematical cuts ({\bf a)} $z>0.25$ and {\bf b)} $z>0.20$). We also include data from the SMC\cite{1,11} experiment. 

\setlength{\unitlength}{1.mm}
\begin{figure}[hbt]
\begin{picture}(120,48)(0,0)
\put(0,0){\mbox{\epsfxsize5.3cm\epsffile{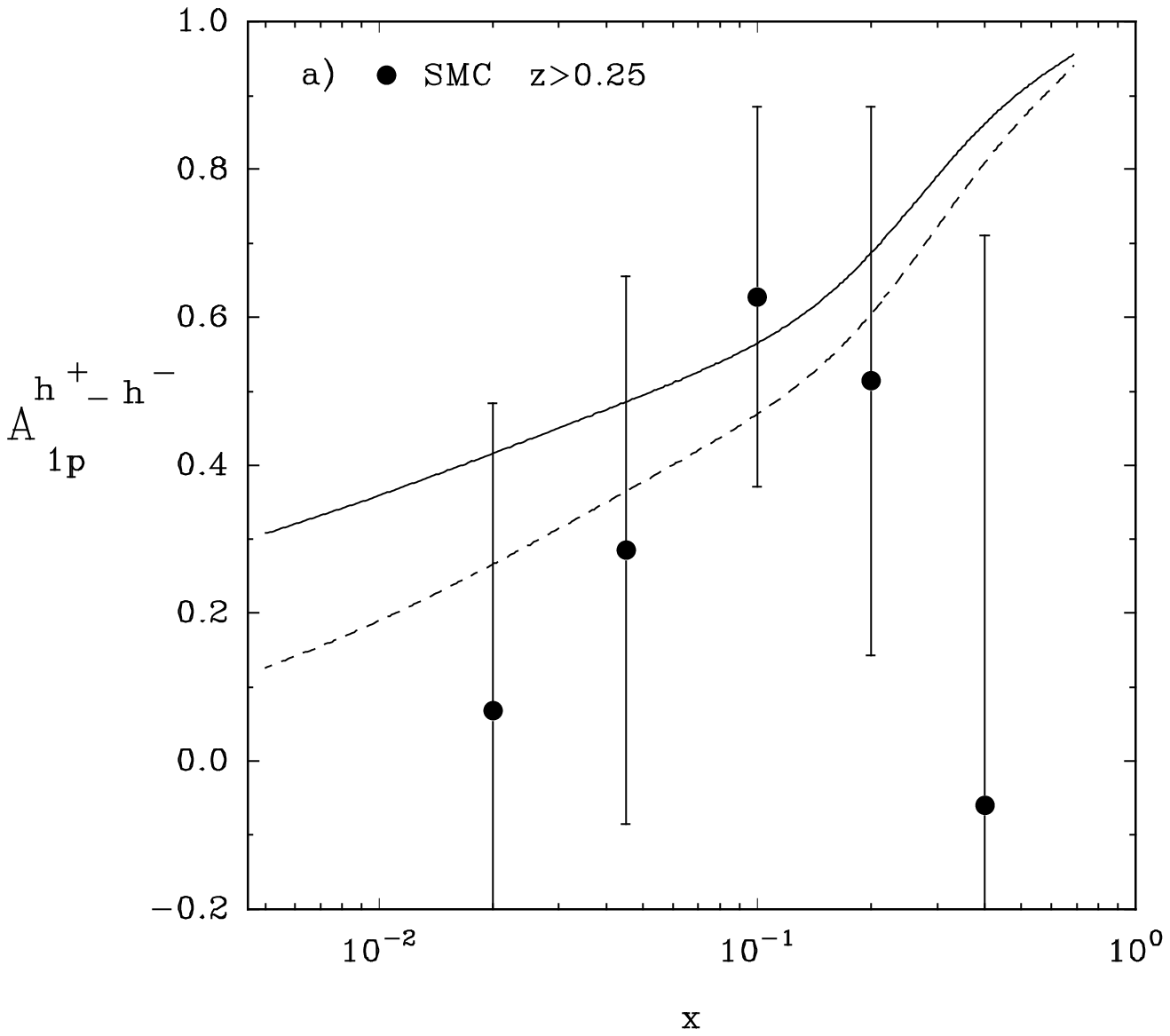}}}
\put(60,0){\mbox{\epsfxsize5.3cm\epsffile{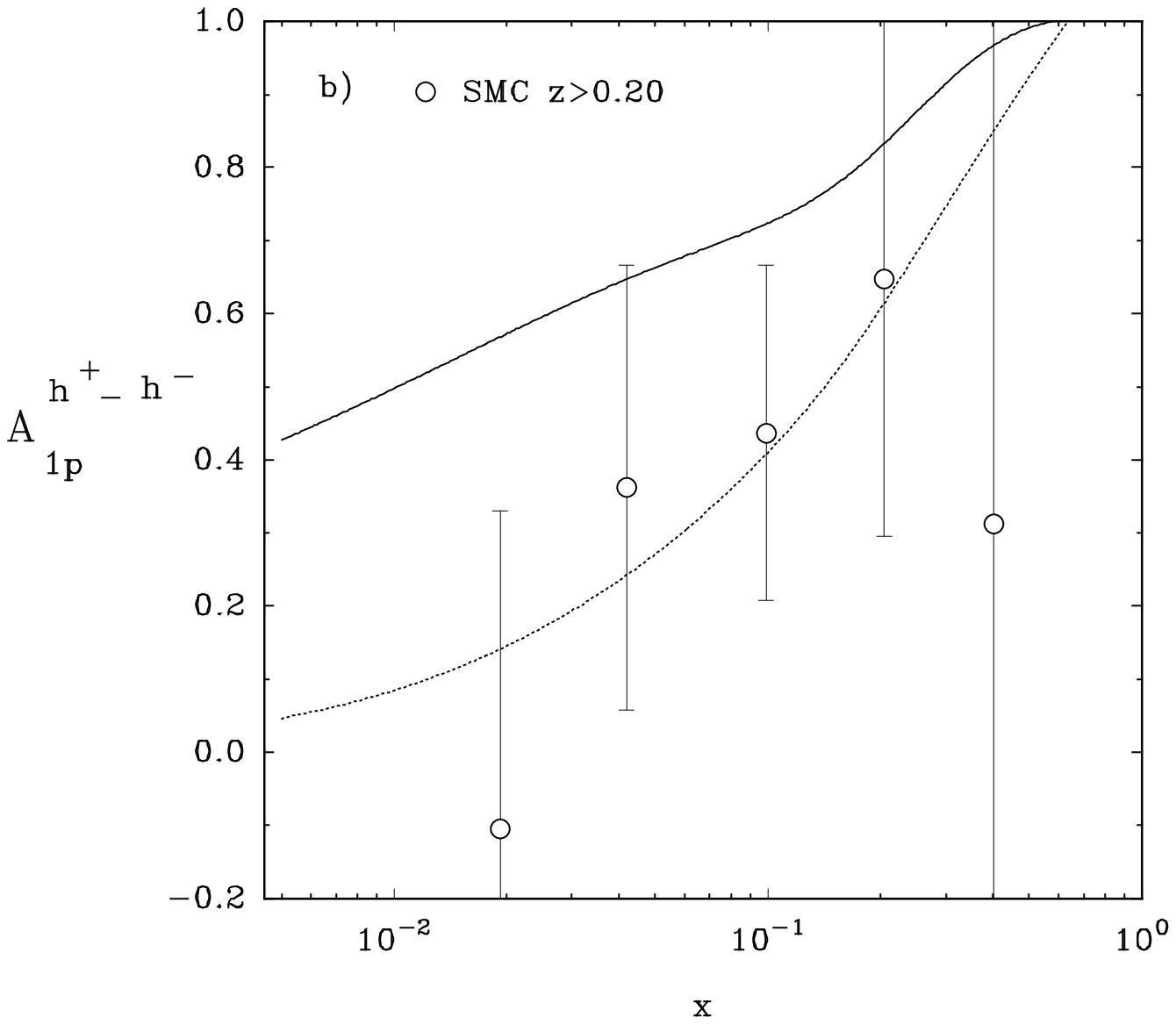}}}
\end{picture}
\caption{Difference asymmetries for:  {\bf a)}  $z>0.25$, {\bf b)}  $z>0.20$
\label{fig:radish}}
\end{figure}

Notice that
at variance with what is found for single asymmetries, here the $z$-cut dependence is much more apparent. For $z>0.10$, figure 3, one finds that the corrections differ in shape and sign, and for the deuteron becomes singular at $z\sim 0.20$.

\setlength{\unitlength}{1.mm}
\begin{figure}[hbt]
\begin{picture}(120,47)(0,0)
\put(30,0){\mbox{\epsfxsize5.3cm\epsffile{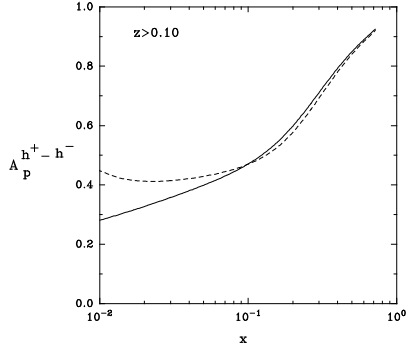}}}
\end{picture}
\caption{The same as figure (2) for  $z>0.10$ 
\label{fig:radish}}
\end{figure}

The unexpected behauvior of these corrections, which are dominated by target fragmentation, is due to the fact that these effects spoil the cancelation of difference between fragmentation functions in expressions like eq.3. These differences vanish in the neighbourhood of $z\sim 0.20$ for pions, making the observable unstable, and for example leading to the singular behauvior in the deuteron asymmetry.

\section*{Conclusions}

Imposing different kinematical cuts, such as those proposed by the COMPASS and the HERMES collaborations, we have computed predictions for both experiments,
using NLO fragmentation functions, parton distributions and also a model for spin dependent fracture functions.

 We have found for the so called ``difference" asymmetries that the corresponding corrections produce significant effects which depend on the inclusion of events where the hadron energy fraction is smaller or greater than approximattely 0.2 units. The value chosen for the $z$-cut not only modifies the size of the corrections but also the sign of them.
 These behaviours are closely related to the fact that pions of opposite charge are produced with almost equal probability in this kinematical region. 

Our estimates imply for single asymmetries that the corrections are almost in the accuracy limits of the forthcoming experiments and that the different choices for the $z$-cuts have no significant consequences in them. However, for
difference asymmetries, the corrections and its dependence on $z$ are crucial for the interpretation of the data and the comparison of the experimental results.

\end{document}